\documentclass[10pt, conference]{IEEEtran}
\IEEEoverridecommandlockouts
\usepackage{cite}
\usepackage{amsmath,amssymb,amsfonts}
\usepackage{algorithmic}
\usepackage{graphicx}
\usepackage{textcomp}

\usepackage{footnote}
\usepackage{multirow}
\usepackage{multicol}
\usepackage{tabulary}

\usepackage{xcolor}
\def\BibTeX{{\rm B\kern-.05em{\sc i\kern-.025em b}\kern-.08em
    T\kern-.1667em\lower.7ex\hbox{E}\kern-.125emX}}
\begin{document}

\title{Semantics Preserving Hierarchy based Retrieval of Indian heritage monuments}

\author{\IEEEauthorblockN{Ronak Gupta \IEEEauthorrefmark{1},  Prerana Mukherjee\IEEEauthorrefmark{2}, Brejesh Lall\IEEEauthorrefmark{1},Varshul Gupta \IEEEauthorrefmark{3}}
\IEEEauthorblockA{\textit{\IEEEauthorrefmark{1}Department of Electrical Engineering, Indian Institute of Technology, Delhi, India}}
\IEEEauthorblockA{\textit{\IEEEauthorrefmark{2}Department of Computer Science, School of Engineering, Jawaharlal Nehru University, India}}
\IEEEauthorblockA{\textit{\IEEEauthorrefmark{3}Department of Computer Science, Jaypee Institute of Information Technology, JDelhi, India}}
 \{ronakgupta143@gmail.com, prerana@jnu.ac.in, brejesh@ee.iitd.ac.in, varshul.cw@gmail.com \} }

% \author{\IEEEauthorblockN{ Prerana Mukherji}
% \IEEEauthorblockA{\textit{Assistant Professor at IIIT Sricity} \\
% \textit{name of organization (of Aff.)}\\
% City, India \\
% email address}
% \and
% \IEEEauthorblockN{ A N Vaishnavi }
% \IEEEauthorblockA{\textit{Undergraduate student at IIIT Sricity} \\
% \textit{name of organization (of Aff.)}\\
% Hyderabad, India \\
% vaish9699@gmail.com}
% \and
% \IEEEauthorblockN{ Gattineni Sai Sreenithya}
% \IEEEauthorblockA{\textit{Undergraduate student at IIIT Sricity} \\
% \textit{name of organization (of Aff.)}\\
% Hyderabad, India \\
% email address}
% \and

% \IEEEauthorblockN{ Gullapalli Keerti}
% \IEEEauthorblockA{\textit{Undergraduate Student at IIIT Sricity} \\
% \textit{name of organization (of Aff.)}\\
% City, India \\
% email address}
% \and
% \IEEEauthorblockN{A Sree Vidya}
% \IEEEauthorblockA{\textit{Undergraduate student at IIIT Sricity} \\
% \textit{name of organization (of Aff.)}\\
% Hyderabad, India \\
% email address}
% \and
% \IEEEauthorblockN{ Deeksha Nayab}
% \IEEEauthorblockA{\textit{Undergraduate student at IIIT Sricity} \\
% \textit{name of organization (of Aff.)}\\
% Hyderabad, India \\
% email address}
% }

\maketitle

\begin{abstract}
Monument classification can be performed on the basis of their appearance and shape from coarse to fine categories. Although there is much semantic information present in the monuments which is reflected in the eras they were built, its type or purpose, the dynasty which established it, etc. Particularly, Indian subcontinent exhibits a huge deal of variation in terms of architectural styles owing to its rich cultural heritage. In this paper, we propose a framework that utilizes hierarchy to preserve semantic information while performing image classification or image retrieval. We encode the learnt deep embeddings to construct a dictionary of images and then utilize a re-ranking framework on the the retrieved results using DeLF features. The semantic information preserved in these embeddings helps to classify unknown monuments at higher level of granularity in hierarchy. We have curated a large, novel Indian heritage monuments dataset comprising of images of historical, cultural and religious importance with subtypes of eras, dynasties and architectural styles. We demonstrate the performance of the proposed framework in image classification and retrieval tasks and compare it with other competing methods on this dataset.

% In ever changing world the attempt is to address one of the significant issues that algorithmically made music had (and still has): the absence of worldwide rationality or structure. 
% In this paper, we will demonstrate our way to deal with producing old style music with rehashed melodic structures utilizing a Long Short Term Memory (LSTM) Neural Network with Attention.

% Attention is a mechanism combined  in the RNN enabling it to concentrate on specific pieces of the information arrangement while anticipating a specific piece of the yield grouping, empowering simpler learning and of higher caliber.

% We proposed the utilization of LSTMs with attention â€” which are evidently superior to vanilla-Recurrent neural network(RNNs) and LSTM for adapting longer fleeting conditions.Because of this experimentation, music has been one of the early utilizations of LSTMs.

%  In this way, with enough information and the right calculation, AI ought to have the option to make music that would sound human. This report plots different ways to deal with music creation through Neural Network models, and despite the fact that there were some blended outcomes by the model, it is clear that melodic thoughts can be gathered from these calculations in order to make another bit of music. 

\end{abstract}


\section{Introduction}
\label{introduction}
Monuments are highly complex three dimensional structures which are constructed to commemorate a person or an event and become an integral part of the cultural heritage due to the historical, political, artistic or architectural significance \cite{goel2012buildings}. In \cite{goel2012buildings}, authors suggested that the buildings and monuments cannot be classified at instance level alone as they have more fine-grained information that can be categorized based on their architectural styles. Architectural styles may not only vary over time but also over geographical regions. In order to discover the characteristic feature across the architectural styles there are several challenges to address the variations in: i) scale, ii) appearance and iii) viewpoint and occlusion. Thus, the discriminative patches to be mined across multiple scales need to be semantically meaningful as well. In \cite{chaudhury2015multimedia}, different applications on multimedia ontology and semantic representations have been elucidated which includes instance retrieval, content based recommendation engines, and information integration. It presents the usage of Multimedia Web Ontology Language (MOWL) to preserve tangible and intangible heritage aspects by leveraging digital heritage. Another seminal work in monuments classification is done by \cite{doersch2015makes}. The authors investigated the geographically relevant features in the visual context such as windows with railings, balcony styles, doorway styles, blue/green/white street signs to classify the Paris buildings. An amalgamation of visual and semantic information can aid in the monument classification and retrieval tasks. This further may guide the application area of interactive virtual tour of heritage sites \cite{chelaramani2017interactive}.

\begin{figure*}[hbtp]
\centering
{
\includegraphics[scale=0.5]{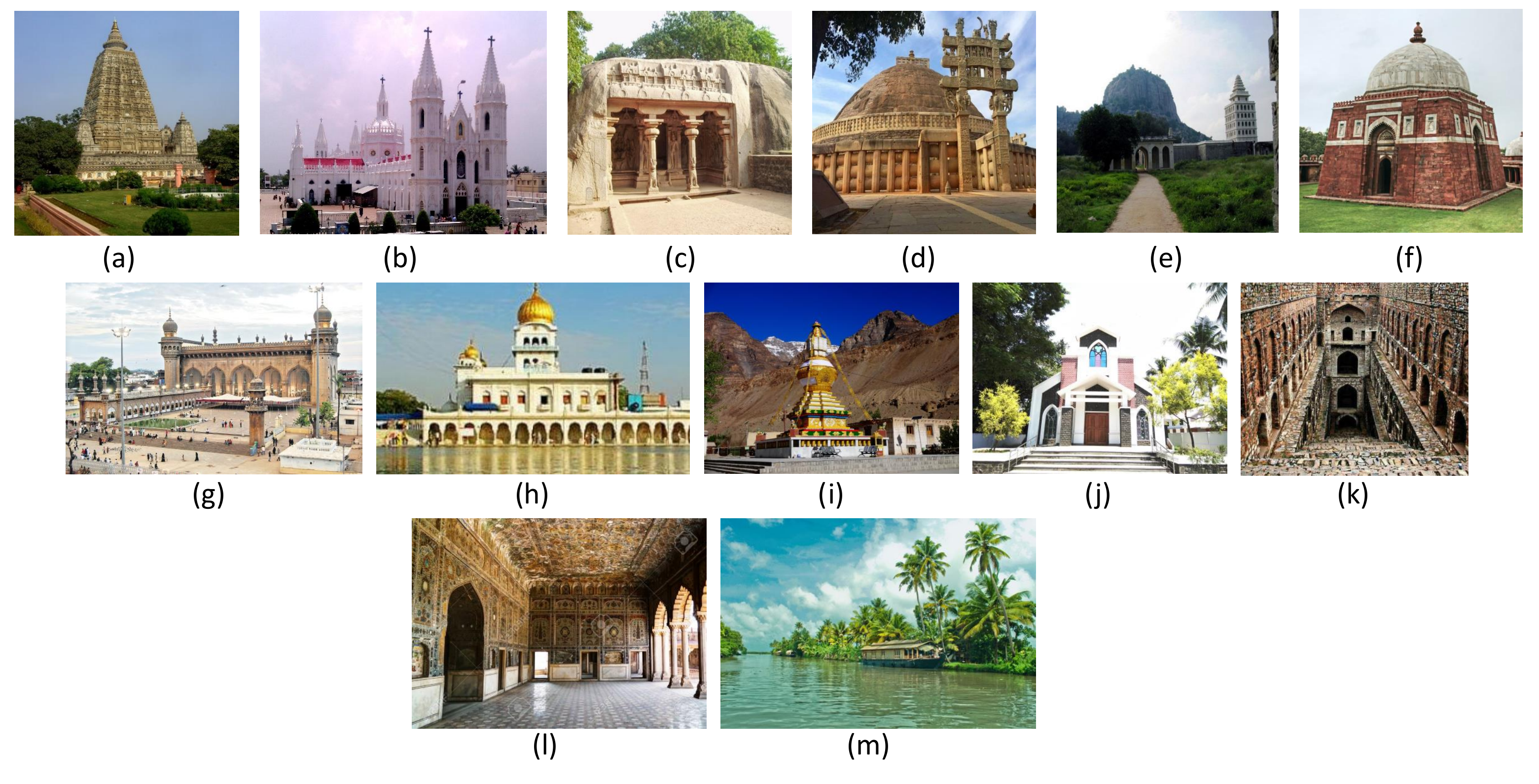}
\caption{Instances of different architectural types: (a) Temple (Mahabodhi temple) (b) Church (Basillica of Valenkanni) (c) Cave (Vsvaraha Cave) (d) Stupa (Sanchi stupa) (e) Fort (Gingee Fort) (f) Tomb (Tomb of Ghiyasuddin Tughlaq) (g) Mosque (Mecca Masjid) (h) Gurudwara (Guru ka Taal) (i) Monastry (Tabo Monastery) (j) Cemetry (Santa Cruz - Cemetry) (k) Baoli (Agrasen Ki Baoli) (l) Palace (Sheesh Mahal) (m) Landscape (Alappuzha).}
\label{fig:instance}
}
\end{figure*}

In this work, we explore the finer nuances in the Indian heritage monument types. India attributes its rich cultural heritage to the reigns of different dynasties belonging to different eras \cite{shukla2017computer}. Even the architectural style across the country is quite varied. Further, due to the variations in the architectural types they can be classified into temples, mosques, gurudwaras, church, fort, tomb, caves, monastry, cemetry, baoli, stupas, landscape and palaces. Fig. \ref{fig:instance} demonstrates the categorical examples of these architectural type classes. These monuments of such varied architectural types were constructed across various eras which can be further subdivided into various dynasty reigns. In order to incorporate such huge diversity in any retrieval methods in computer vision, it requires a lot of semantic knowledge embedded in the learning paradigm. Most of the retrieval methods \cite{radenovic2018fine, xu2018cross, song2018binary, zhao2015deep, lu2016latent} rely upon the instance level search based on the semantic similarity however such architectures do not allow category level drill-down at the same time. 

Recently, a considerable amount of research has been focused towards solving instance based retrieval in multimedia and computer vision. The problem of finding and retrieving similar or exact instances of the query image from a large pool of images refers to content based image retrieval. The focus has been now shifted towards semantic retrieval rather than just restricting it to the domain of instance based retrieval. Vector encoding strategies such as bag of visual words \cite{sivic2003video}, fisher vector \cite{sanchez2013image} etc. has remained the de-facto standard for retrieval pipelines using local features. Once the descriptors are extracted from images, similar images can be then retrieved using inverted file index organisation or hashing based methods. In order to incorporate geometric verification in bag of visual words, \cite{philbin2007object} introduced a variant of bag of visual words. It enables to retrieve multiple view variations of the query image. Due to the representational ability of deep convolutional neural networks (CNNs) \cite{krizhevsky2012imagenet} as they learn rich mid-level image descriptors, feature vectors from last layers have shown exemplary performance in image retrieval as well. To reduce the feature dimensionality for matching purposes, PCA based techniques are further augmented. In \cite{lin2015deep}, authors utilize deep CNNs to construct hash codes to speed up retrieval. 
Archaeological experts with the domain knowledge can easily categorize monuments based on the given characteristics. 

Automating the task of identifying the fine-grained classes in case of monuments and infusing semantic information of the hierarchy in the feature embedding given no a-priori knowledge is quite daunting task. In this paper, we construct a hierarchy tree of various eras subdivided into different architectural styles further classified into dynasties as the leaf node. We explore the structural nuances in such architectural styles so that features which are representative of the particular era can be clustered together. Finally, a dictionary is constructed based on the deep semantic embeddings obtained and re-ranking framework is imposed on top of it utilizing DeLF features \cite{noh2017large}. A query image can belong to classes (monuments) already seen in the training phase or can be an unseen/unknown monument (novel classes) in the retrieval phase. 

Thus, owing to the complexity of variations present in Indian monuments due to its rich cultural influence, we primarily focus to propose a framework that is capable of describing holistic attributes in the retrieval mechanism particularly in context of Indian heritage monuments. In this paper, we have tried to address the following questions:
\begin{itemize}
    \item What type of fine-grained information can we infer from images of Indian monuments-particularly eras, dynasty reign and architectural styles?
    \item Can semantically learnt deep representations be utilized for constructing a unified framework capable of predicting multiple attributes for a given image of an Indian monument and consequently strengthen the retrieval pipeline? 
\end{itemize}
In view of above discussions, the key contributions can be summarized as
\begin{enumerate}
\item To the best of our knowledge, this is the first work to holistically infuse semantic hierarchy preserving embeddings to learn deep representations for retrieval in the domain of Indian heritage monuments. 
\item We encode the learnt embeddings to construct a dictionary of images and then utilize a re-ranking framework on the retrieval results using DeLF features. 
\item We show the applicability of proposed framework for fine-grained classification and semantics preserving retrieval.
\item We have curated a novel data-set comprising of monument images of historical, cultural and religious importance from all over India namely Indian heritage monuments dataset (IHMD). These monuments are further subdivided into different eras, dynasties and architectural styles. 
\end{enumerate}
Overall architecture of the paper is as follows. Related works is given in Section \ref{sec:related}. The proposed methodology is explained in Section \ref{sec:methodology}. Experimental results and discussion are given in Section \ref{sec:results} followed by conclusion in Section \ref{sec:conclusion}.
%%%%%%%%%%%%%%%%%%%%%%%%%%%%%%%%%%%%%%%%%%%%%%%%%%%%%%%%%%%%%%%%%%%%%%%%%%
\section{Related work}
\label{sec:related}
\subsection{Handcrafted features based retrieval}
In \cite{johnson2015image}, authors developed a framework for image retrieval using scene graphs. Conditional Random Fields (CRFs) are utilized to map the objects, attributes and relationships between them into the scene graph. Likelihood scores based on CRF graphs are taken into account for retrieval. In \cite{philbin2007object}, authors performed retrieval on large scale dataset utilizing spatial verification in the ranking process and introduced a novel quantization technique using randomized trees. In \cite{csurka2004visual}, authors pioneered the bag of keypoint concept where vector quantization methods were utilized to encode the affine invariant descriptors on image patches. The de-facto standard for feature detectors Scale Invariant Feature Transform (SIFT) and Speeded Up Robust Features (SURF) was championed by ORB \cite{rublee2011orb} which utilized binary descriptor named BRIEF and was invariant to rotation and noise variations. ORB has been predominantly utilized in feature matching applications thereafter. In \cite{arandjelovic2012three}, authors introduced a new robust descriptor RootSIFT
and novel method for query expansion to discriminatively learn the query features with efficient use of inverted file index structures. Various feature quantization methods like Fisher Vector \cite{sanchez2013image}, VLAD \cite{delhumeau2013revisiting} also gained popularity for image retrieval tasks. 

Another set of works are in the domain of content based image retrieval (CBIR). In \cite{xia2016privacy}, CBIR based techniques are used to store and retrieve images over cloud server. First, feature vectors are extracted and then locality sensitive hashing is used to construct pre filter table. Finally, using secure K-nearest neighbor pixels are encrypted with secure cipher. In \cite{zhu2016unsupervised}, the rich semantic information embedded in the text helps in boosting the performance of unsupervised visual hashing of images. In \cite{piras2017information}, a detailed review of the CBIR techniques has been elucidated. In \cite{revaud2019learning}, author presented an ensemble network of deep CNN based and low level features like color or texture constructed by vector quantization techniques to perform the task of content based image retrieval.

\subsection{Deep features based retrieval}
In \cite{babenko2015aggregating}, authors investigated aggregation of local deep descriptors to produce a global descriptor for efficient image retrieval. In \cite{gordo2016deep}, authors generated region wise descriptors and aggregated to form a global descriptor for instance based search. Thus, region proposal network is incorporated to select the regions which should be the constituent of the global descriptor followed by learning a triplet loss to segregate the dissimilar pairs from similar pairs. In \cite{gordo2017end}, authors created a large scale noisy Landmarks dataset for image retrieval tasks and also show improvements on the global deep descriptor based on Siamese architecture. In \cite{salvador2016faster}, authors proposed an instance search pipeline based on object proposal networks such as Faster-RCNN and leverage the deep features followed by a spatial re-ranking framework. In \cite{tolias2015particular}, authors utilized integral images to handle max pooling in the convolutional layer activations. In \cite{altwaijry2016learning}, authors utilized deep correspondence matching based on image patches in aerial images for retrieval task. In \cite{teichmann2018detect}, authors utilized a novel regional aggregated selective match kernel to combine the information from the regions detected into a holistic description. In \cite{kim2018regional}, authors augmented regional-attention network with Regional-Maximum Activation of Convolutions (R-MAC) to improve the retrieval performance. In \cite{lin2015deep}, authors utilized deep supervised hashing to learn compact similarity preserving binary codes for fast and efficient retrieval. They discriminatively learn the similarity between the pair of images and produce binary hash codes in a manner that similar images are mapped closer and vice-versa for the dissimilar images. In \cite{zhao2015deep}, authors extended the binary hash code to multilabel semantic image retrieval. In this work, the authors construct hash codes based on deep CNNs which jointly learns the feature encoding and mappings based on multilevel semantic ranking. 

% \subsection{Hashcode based retrieval}
% \cite{lin2015deep} utilized deep supervised hashing to learn compact similarity preserving binary codes for fast and efficient retrieval. They discriminatively learn the similarity between the pair of images and produce binary hash codes in a manner that similar images are mapped closer and vice-versa for the dissimilar images. \cite{zhao2015deep} extended the binary hash code to multilabel semantic image retrieval. In this work, the authors construct hash codes based on deep CNNs which jointly learns the feature encoding and mappings based on multilevel semantic ranking. \cite{bai2018adaptive} introduce novel hash codes based on kernel function to have a similarity search coherent to k-nearest neighbor approach. \cite{guo2017learning} utilize sparse hashing based methods for retrieval and optimize the anchor sets in sparse representations. \cite{shen2016semi} introduce an unsupervised hashing approach based on semi-paired cross view retrieval. \cite{wang2017few} focus on few shot hash learning to learn the universal hash functions offline from a pool of unlabeled data. 

%%%%%%%%%%%%%%%%%%%%%%%%%%%%%%%%%%%%%%%%%%%%%%%%%%%%%%%%%%%%%%%%%%%%%%%%%%

\section{Proposed Methodology}
\label{sec:methodology}
\begin{figure*}[hbtp]
\centering

$\begin{array}{cc}
\centering
(a) &
\includegraphics[width=17cm,height=6cm]{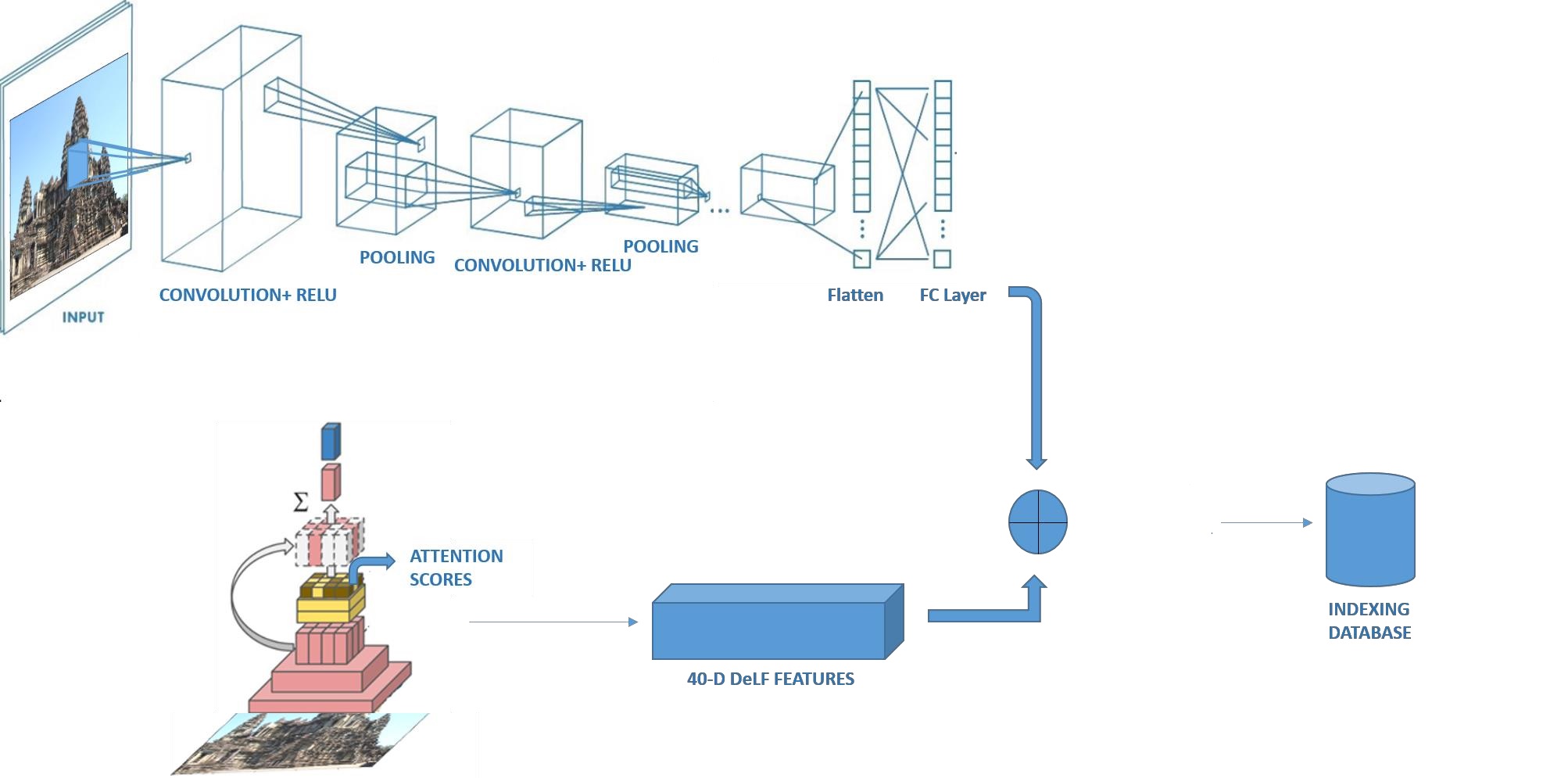} \\
(b) &
\includegraphics[width=15cm,height=6cm]{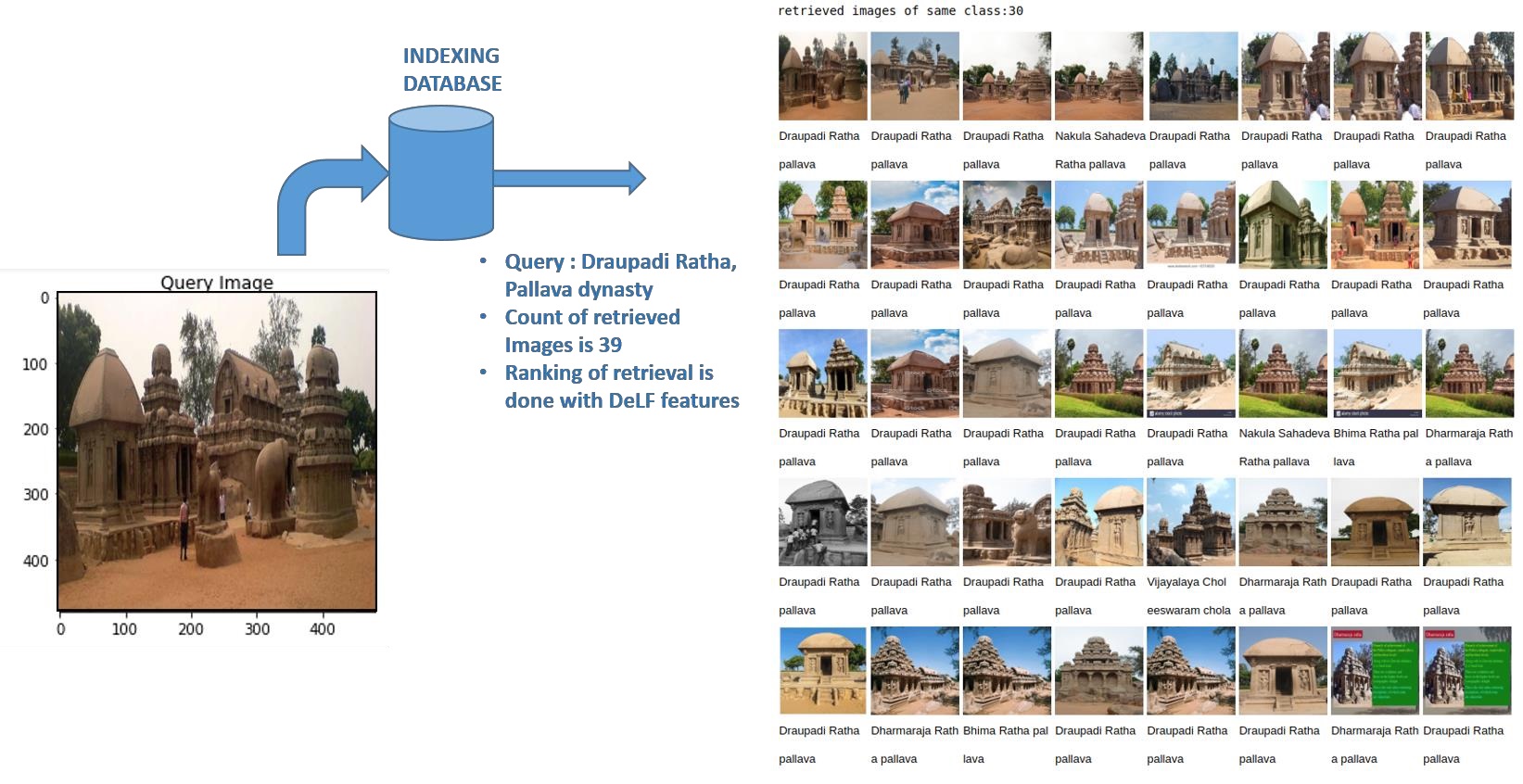} \\

\end{array}$
\caption{a) Proposed Pipeline for indexing images b) Retrieval of Top-N images based on query image search with re-ranking. First image in the retrieval set depicts the query image itself}
\label{fig:workflow}
\end{figure*}

%\vspace{-0.2cm}
In this section, we give a detailed overview of the proposed semantic hierarchy preserving retrieval scheme. The end-to-end pipeline of the proposed method is shown in Fig. \ref{fig:workflow}. % In the following subsections, we describe the components of the proposed method in detail.

% \begin{figure*}[thb]
%       \centering
%       \fbox{
%       \includegraphics[scale=0.7]{Proposed.pdf}}
% \caption{a) Proposed Pipeline for indexing images b) Retrieval of Top-N images based on query image search with re-ranking. First image in the retrieval set depicts the query image itself.}
% \label{fig:workflow}
%       %\vspace{-0.25in}
% \end{figure*}

\subsection{Semantic Feature Embedding}
\label{sec:edgesaliency}
%\vspace{-0.2cm}
Most of the existing work in context of retrieval has been done in content based image retrieval (CBIR). Recently, incorporating semantics into retrieval is gaining attention. However, infusing this information only at the class-level may not rigorously embed the parent-child relationships. Learning this embedding at cross-class level is even difficult. Therefore, in order to deal with this we capture domain knowledge in the form of a hierarchy tree to aid the retrieval pipeline. In \cite{barz2019hierarchy}, authors introduced hierarchy tree concept for generic object dataset CIFAR-100 where the class-level intricacies are less. We get inspired from \cite{barz2019hierarchy}, and take it a level-forward in case of monuments and construct a hierarchy tree with careful selection of nodes. However, in contrast to \cite{barz2019hierarchy} which utilizes prior knowledge only for semantic image retrieval in generic object datasets, in our proposed framework, we jointly learn semantic embeddings such that all the class embeddings lie on a unit hypersphere and visual feature embeddings such that the attention function when applied to the last fully convolutional layer of CNN (DeLF features, as described in Sec. \ref{sec:delf}) gives us a global descriptor for each image. For e.g., the semantic embeddings for two different tombs will lie close on the hypersphere. The objective of semantics preserving image retrieval for our Indian heritage dataset with the given semantic relations has not been presented earlier in this domain. 

% The semantic embedding models in our framework are trained for novel Indian heritage monuments dataset as curated for this work.
% The correlation of the semantic embeddings between given classes will be equal to semantic similarity of classes defined by the lowest common subsumer of the given classes.  

The reason why Indian monuments are challenging are as follows:
\begin{itemize}
    \item Indian monuments show huge variations across eras, dynasties, architectural styles. However, different architectural styles within same dynasties may have similarities. Similarly, within same eras, different architectural styles may be built in same fashion.
    This requires not only similarity by part but also a global descriptor which inherently infuses such distribution of semantic embedding space.
    \item Cross-cultural architectural monuments may also exhibit strong data association. 
    \item  In CBIR based techniques, the data which is visually similar in terms of cues such as color, texture, structure may get clustered together even though they are not semantically linked. For e.g. in case of CBIR based techniques, Jama Masjid can be strong match for Taj Mahal although semantically both are not related. Jama Masjid is for religious prayers while Taj Mahal which is rather more close to Humayun's Tomb as they both fall in the broad category of tombs. Thus, visual as well as semantic similarity both play a strong cue in retrieval.
\end{itemize}
Most of the traditional semantic learning approaches (\cite{zhao2011large, chang2015large}) utilize class level knowledge in classification or retrieval pipeline at the classifier level or utilize taxonomy embeddings however they do not semantically learn the feature embedding space as done in \cite{barz2019hierarchy}. It helps in integrating relevance feedback in the sequence. Only visual embeddings are not enough to map this taxonomic relevance to the retrieval engine rather word embeddings need to be incorporated as well. Instead of training on huge corpus of text, rather utilizing class embeddings based on prior knowledge derived from hierarchy tree of classes is more reliable and less complex.

% Motivated by these intuitions, we however try to have an ensemble of visual and semantic learning for retrieving data. 
% In order to learn such overt and covert similarities, we construct the semantic embedding space influenced by the hierarchy tree where such features which are semantically similar will be close and those which are quite distinct fall apart. Since, the convolutional neural network based features act as global descriptors, we further couple this with local descriptors for visual similarity search. For this, we used pre-trained DeLF model and use it to generate 40-dimensional vector representing local features in the image.

\begin{figure*}[hbtp]
\centering
{
\includegraphics[scale=0.3]{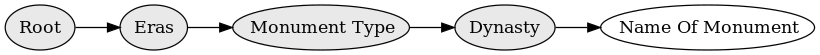}
\caption{Levels of hierarchy tree for Indian heritage monuments dataset}
\label{fig:read}
}
\end{figure*}

\begin{figure*}[hbtp]
\centering
{
\includegraphics[width=12cm]{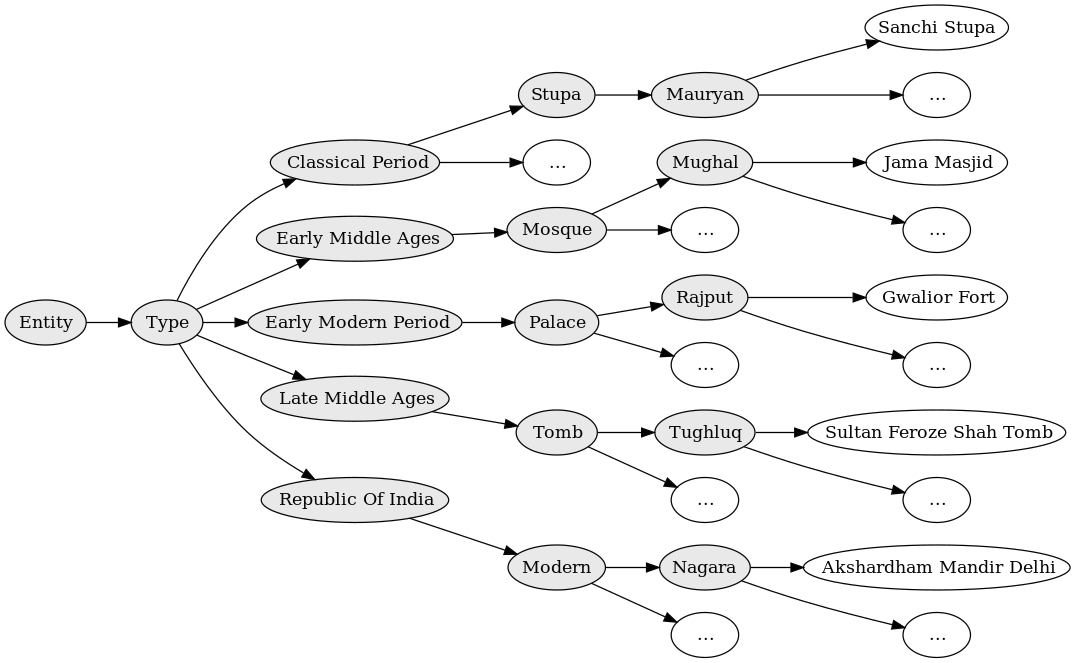}
\caption{Toy hierarchy tree for Indian heritage monuments dataset}
\label{fig:hierarchy}
}
\end{figure*}

For a directed acyclic graph $\mathcal{G(V,E)}$, where $\mathcal{V}$ denotes the nodes and $\mathcal{E}$, denotes the edges specifying the hyponymy relation between semantic concepts. For Indian heritage monuments dataset, we construct extensive hyponym and hypernym creation along with 4-level tree for better domain representation. We maintain a trade-off between region level annotation and ontology infusion. Edge $(u,v)$ would mean that $u$ is a superclass of $v$. Classes of interest are denoted by $\mathcal{C}=\{c_1,...,c_n\}\subseteq \mathcal{V}$ are a subset of semantic concepts. The dissimilarity between two classes is given as the height of the hierarchy tree which is further calculated using the \textit{lowest common subsumer (LCS)} of two classes as,
\begin{equation}
    d_\mathcal{G}(u,v)=\frac{height(LCS(u,v)}{max_{w\epsilon \mathcal{V}} height(w)}
\end{equation}
where the height of a node is the depth of the tree from that node to leaf. The LCS of two nodes is the ancestor of both nodes that does not have any child
being an ancestor of both nodes as well.

Thus, the semantic similarity [0-1] between two classes is given as,
\begin{equation}
    s_\mathcal{G}(u,v)=1-d_\mathcal{G}(u,v)
\end{equation}
The next step is to compute the class embedding $\phi(\mathcal{C}_i)\epsilon \mathbb{R}^n $, $\forall c_i \epsilon \mathcal{C}$ such that,
\begin{align}
  \forall_{1\leqslant i,j  \leqslant n} : \phi (c_i)^T \phi (c_j)=s_\mathcal{G}(u,v) \\
  \forall_{1\leqslant i  \leqslant n} : \left \| \phi (c_i) \right \|=1
\end{align}
where all the classes lie on a unit hypersphere, if a new embedding is present the dot product with the already embedded class will give the semantic similarity of the two classes.
After the target embeddings for all classes is precomputed, we need a transformation $\psi: X\rightarrow \mathbb{R}^n$ from the image space into
 hierarchy-based semantic embedding space,
so that image features are close to the centroid of their class. This transformation or feature representation is learned as presented in next section \ref{sec:learning}.

\subsection{Learning Semantic Embeddings using CNNs}
\label{sec:learning}
We employ CNNs to learn the feature representation that correlates to semantic embedding of the class. The correlation loss as defined in \cite{barz2019hierarchy} enforces similarity between the learned image representations and the semantic embedding of the class. Usually, the features extracted from the last layer of the same CNN network architecture, trained with cross-entropy loss gives a good feature embedding. Hence when adding a fully-connected layer with softmax function on top of the embedding layer, the network can learn simultaneously semantic image embeddings and classify images based on combined loss.
% We train a correlation loss function that complies similarity between the learned image representations and the semantic embedding of their class in congruent lines to \cite{barz2019hierarchy}. Thus, the images are then stored in the indexed database with semantic similarity embeddings learnt.

\subsection{DeLF Feature based Reranking}
\label{sec:delf}
Next, in order to embed visual similarity further to the semantic similarity in the retrieval pipeline we further re-rank the retrieved images with DeLF based features \cite{noh2017large}. For a given image, dense features are extracted using a fully convolutional network (FCN) by training with a classification loss. We utilized pretrained ResNet50 architecture \cite{he2016deep} on ImageNet \cite{russakovsky2015imagenet} for obtaining the FCN features. The weights are finetuned on Google Landmarks dataset \cite{noh2017large}. In order to handle the scale variations, pyramidal network is employed. Following that, an attention layer is incorporated to select the subset of features. The attention score function $\alpha(.)$ is designed with a 2-layer CNN followed by softplus activation function. Let $f_n\epsilon \mathbb{R}^n, n=1,...,N$ denote the d-dimentional features to be jointly learnt with attention model. An attention score function $\alpha(\mathbf{f}_n;\theta)$ is learnt for each feature, where $\theta$ denotes the parameters of function $\alpha(.)$. The output logit $\bf{y}$ is obtained as, 
\begin{equation}
    \mathbf{y}=\mathbf{W}\sum_n \alpha((\mathbf{f}_n;\theta).\mathbf{f}_n)
\end{equation}
where $\mathbf{W}\epsilon \mathbb{R}^{Mxd}$ denotes the weights of the final FC layer of the CNN trained to predict $M$  classes. For training, cross entropy loss is utilized as in \cite{noh2017large}. Thus, we do not need to describe the feature points as we get a global descriptor of dimensionality 1$\times$40 for each image. This enables to incorporate visual similarity for retrieval of images.

\subsection{Retrieval}
\label{sec:winscore}
%\vspace{-0.2cm}
Since, the CBIR based techniques just take into account visual similarity thus, mean average precision (mAP) and precision @ top-$K$ results are evaluated. In order to incorporate the semantic similarity in the retrieval, we require a heirarchical precision metric ($HP@k, 1\leqslant K \leqslant m$) as adopted from \cite{deng2011hierarchical}. It is defined as,
\begin{equation}
    HP@k(R)=\frac{\sum_i=1^k s_\mathcal{G}(y_q,y_i)}{max_\pi \sum_i=1^k s_\mathcal{G}(y_q,y_{\pi_i})}
\end{equation}
where $\pi$ denotes the permutation of indices from $1$ to $m$. Denominator helps in normalizing the sum of precisions. Here, $x_q \epsilon \mathcal{X}$ denotes the query image with class label $y_q \epsilon \mathcal{Y}={1,...,n}$ and retrieved images are denoted by $\mathcal{R}=(x_1,y_1), ..., (x_m,y_m)$, where each retrieved image is an ordered pair of image $x_i$ and its associated label $y_i$. The area under the curve for $k$ ranging from $1$ to $K$ denotes the average hierarchical precision @K. Thus, the semantic similarity can be measured for the retrieved set of images. 
%%%%%%%%%%%%%%%%%%%%%%%%%%%%%%%%%%%%%%%%%%%%%%%%%%%%%%%%%%%%%%%%%%%%%%%%%%

\section{Experimental Results}
\label{sec:results}
%\vspace{-0.2cm}

\subsection{Experimental Setup and Parameter Settings}
We perform the experiments in similar lines to \cite{barz2019hierarchy} on the curated Indian Heritage Monuments dataset. The efficacy of the semantic embeddings on generic object dataset such as CIFAR-100 has been validated in \cite{barz2019hierarchy}. In order to make our framework applicable to Indian heritage monuments dataset, we created a taxonomy for the set of classes present in this dataset. A toy hierarchy tree which is subset of the full taxonomy is shown in Fig. \ref{fig:hierarchy}. There is a strict process of branching out at different levels of the taxonomy as shown in Fig. ~\ref{fig:read}. Different semantic relations are captured at different levels of the taxonomy. First level is the era in which the monument is established, the second level is the monument type which represents the purpose of the establishment, the third level is the dynasty under which the monument is established which captures the architecture styles and finally the fourth level is the name of the monument itself.

\begin{table}[hbtp]
\footnotesize
\centering
\begin{tabular}{|l|l|l|l|l|}
\hline
\multicolumn{1}{|c|}{\multirow{2}{*}{\textbf{Method}}}                                                  & \multicolumn{4}{c|}{\textbf{Indian Heritage Monuments Dataset}}                                           \\ \cline{2-5} 
\multicolumn{1}{|c|}{}                                                                         & \multicolumn{1}{c|}{Plain-11} & \multicolumn{1}{c|}{ResNet-34} & ResNet-50 & ResNet-50* \\ \hline
Classification-based                                                                           & 0.6121                        & 0.5262                         & 0.5212    & 0.5903     \\ \hline
Label Embedding                                                                                & 0.6153                        & 0.4052                         & 0.5960    & 0.6620     \\ \hline
\textbf{\begin{tabular}[c]{@{}l@{}}Semantic Embeddings +\\ DeLF based re-ranking\end{tabular}} & 0.5481                        & 0.7083                         & 0.8488    & 0.8823    \\ \hline
\end{tabular}
\caption{Retrieval performance of different image features in mAHP@40}
\label{tab:reter_mahp}
\end{table}

We perform training from scratch for learning semantic embeddings with different convolutional neural network architectures and compare its performance with pre-trained CNN architecture on ImageNet after doing transfer learning. We do not rely upon off-the-shelf pre-trained semantic embedding models trained over other datasets. The Indian heritage dataset is very distinct from other generic object datasets like CIFAR-100/ImageNet categories as it contains much finer nuances in terms of architectural styles, therefore utilizing pretrained networks without the prior knowledge of semantic relationships doesn’t improve the mean Average Hierarchical Precision (mAHP). One of the architecture is Plain-11, which is a VGG style network with 11 trainable layers. Other two are vanilla ResNet-34 and ResNet-50, which are regular deep residual networks with 34 and 50 trainable layers. For ResNet-50 network, we also compare results of model with and without pre-training on ImageNet dataset. (*) denotes the network with pre-training. 

\begin{table}[hbtp]
\footnotesize
\centering
\begin{tabular}{|l|l|l|l|l|}
\hline
\multicolumn{1}{|c|}{\multirow{2}{*}{\textbf{Method}}} & \multicolumn{4}{c|}{\textbf{Indian Heritage Monuments Dataset}}                                  \\ \cline{2-5} 
\multicolumn{1}{|c|}{}                                 & \multicolumn{1}{c|}{Plain-11} & \multicolumn{1}{c|}{ResNet-34} & ResNet-50 & ResNet-50* \\ \hline
Classification-based                                   & 66.16\%                       & 67.33\%                        & 67.59\%   & 77.81\%    \\ \hline
Label Embedding                                        & 70.62\%                       & 71.46\%                                & 75.62\%   & 81.04\%    \\ \hline
\textbf{Semantic Embeddings}                           & 61.4\%                        & 67.24\%                        & 72.87\%   & 77.42\%    \\ \hline
\end{tabular}
\caption{Classification accuracy of various methods over Indian heritage monuments dataset}
\label{tab:classacc}
\end{table}

%\\

For the training process of all the networks, we use a cyclical learning rate with Stochastic Gradient Descent with Warm Restarts (SGDR) \cite{loshchilov2016sgdr}. This refreshes the model learning by setting the learning rate as near as zero for every 12, 36 and 84 and 180 epochs. This is primarily used so that the network doesn’t get stuck in a local minima and the model learning is more generalised over the entire dataset. All the networks were trained for 180 epochs with a batch size of 8. In case of pre-trained ResNet-50 model, the fine-tuning over monument dataset is performed for 8 epochs only. In the curated Indian heritage monuments dataset, each class consists of 50 images over 143 monument classes. The train to test ratio split is 8:2.

\subsection{Indian Heritage Monuments dataset details}
% Several monuments from 200 classes had very few relevant samples (less than 10) and consisted of clutter in terms of people standing in front of the monument, occlusion of the facade due to the view, watermark, etc. Thus, after initial filtering the dataset consists of 143 classes where we have a balanced number of samples.

 We collected the data from Google images, Wikimedia and Flickr using webscraping by web APIs\footnote{https://github.com/hardikvasa/google-images-download}. The initial list of monuments were obtained from Archaeological Survey of India (ASI) website\footnote{http://asi.nic.in/} along with Wikipedia. After extracting images for ~200 monuments, we applied DeLF based features to cluster images as a preprocessing step to filter irrelevant and redundant images. Since this is a retrieval problem, we ensured that they have enough intra-class and inter-class variance so that the proposed method addresses issues related to scale, occlussion, low illumination/night condition imagery etc. Several monuments from 200 classes had very few relevant samples (less than 10) and consisted of clutter in terms of people standing in front of the monument, occlusion of the facade due to the view, watermark, etc. Thus, after initial filtering the dataset consists of 143 classes where we have a balanced number of samples with mean sample per class as 50. Once we prepared the Indian Heritage monuments dataset (IHMD) \footnote{IHMD is available for research upon request.}, we created a hierarchy tree manually, under domain knowledge and expertise. The image size is of 480x480. The architectural styles of the monument types included are church, gurudwara, monastery, mosque, palace, stupa, masjid, tomb, fort, temple. The eras included are Classical Period (320 BCE-550 CE), Early Middle Ages (550 CE-1200 CE), Late Middle Ages (1100 CE-1526 CE), Early Modern Period (1500 CE-1858 CE) and Republic of India (1947 CE-present). The dynasties included in the hierarchy tree are Nagara, Tughlaq, Sayyad, Lodi, Khilji, Rajputs, Cholas, Kakatiyas, Mughals, Pandyas, Rashtrakutas, Pallavas, Karkotas, Bhumijas, Shakaras, Pratiharas, Guptas, Mauryans. The hierarchy tree for the Indian heritage monuments dataset is given in Fig. \ref{fig:hierarchy}.

\begin{table}[hbtp]
\footnotesize
\centering
\begin{tabular}{|c|c|c|}
\hline
\textbf{Input size} & \textbf{\begin{tabular}[c]{@{}c@{}}Semantic Embeddings\\ (ResNet 50) Accuracy\end{tabular}} & \textbf{\begin{tabular}[c]{@{}c@{}}Test speed\\ (Images/sec)\end{tabular}} \\ \hline
224x224             & 62.08\%                                                                                     &       242                                                                   \\ \hline
320x320             & 70.3\%                                                                                      & 128                                                                      \\ \hline
480x480             & 72.87\%                                                                                     & 78                                                                       \\ \hline
\end{tabular}
\caption{Trade-off between classification performance and testing speed of semantic embeddings method}
\label{tab:inference}
\end{table}

\begin{table}[hbtp]
\footnotesize
    \centering
    \begin{tabular}{|c|c|c|}
    \hline
    \textbf{Tombs}     &  Sayyid Tomb & Lodi Tomb \\
    \hline
    Sayyid Tomb     &  80\%     & 10 \% \\
    \hline
    Lodi Tomb       & 66 \%  & 10\% \\
    \hline
    \end{tabular}
    \caption{Confusion between Visually Similar sub-classes}
    \label{tab:confused}
\end{table}

There are few other works which have addressed Indian architectural styles. The dataset used by \cite{kumar2018improving} is limited to four architectural styles namely, Buddhist (8 classes), Kalinga (7 classes), Dravidian (8 classes) and Mughal (4 classes) with a total of 3514 images in 27 classes. While our dataset has high semantic information where we cover the type of the monument (architectural style) (11 classes), ruling dynasty when it was built (37 classes) and the monument classes(143 classes) with a total of 7150 images. In IHIRD dataset ~\cite{podder2018ontology}, authors capture several individual structures (or sculptures) found in the monument instead of the facade of the monument as done in our dataset. For e.g. there are images of Nandi bull (sculpture) present in front of Lord Shiva temple.

%%%%%%%%%%%%%%%%%%%%%%%%%%%%%%%%%%%%%%%%%%%%%%%%%%%%%%%%%%%%%%%%%%%%%%%%%%
\begin{figure}[hbtp]
    \centering
    \includegraphics[scale=0.5]{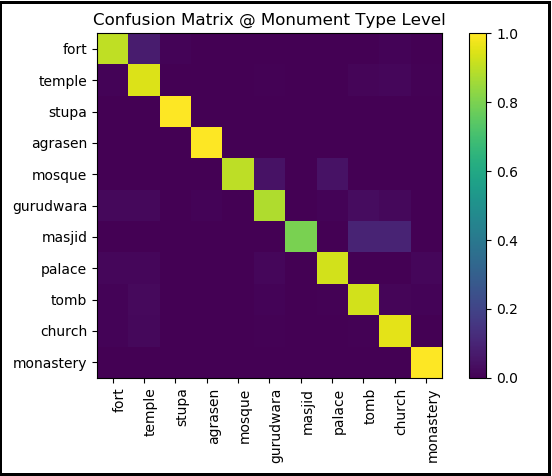}
    \caption{Confusion matrix for monument type classification with Semantic Embeddings from ResNet-50 finetuned network}
    \label{fig:conf_mon_type}
\end{figure}
%\\

\subsection{Quantitative Evaluation}
\begin{figure}
    \centering
    \includegraphics[width=8cm]{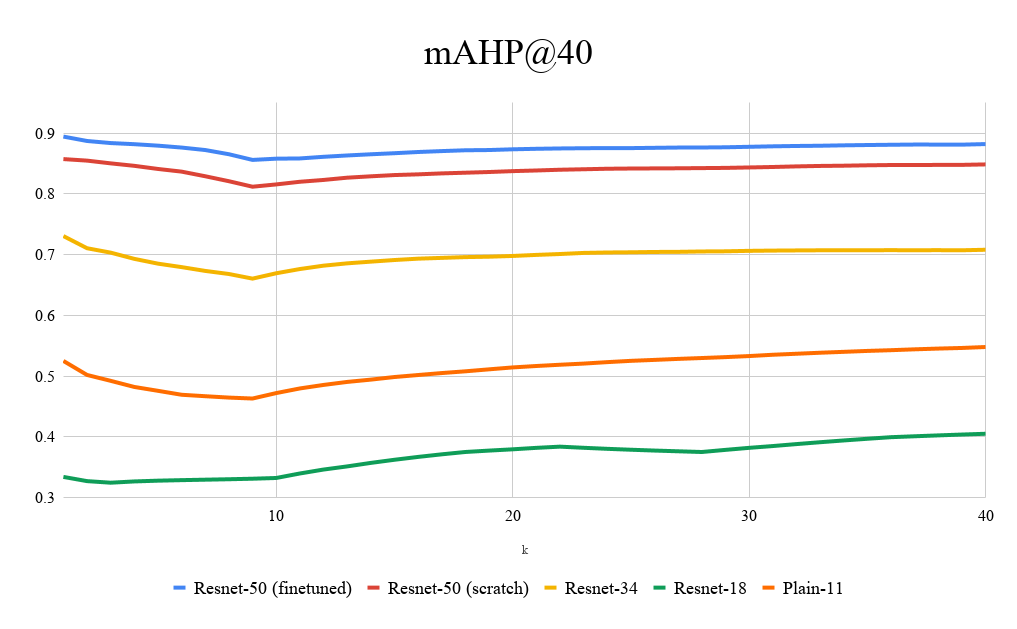}
    \caption{Hierarchical Precision on Indian Heritage Monuments Dataset with different architectures in semantic embeddings}
    \label{fig:mahp40}
\end{figure}

Image retrieval when evaluated using the precision of top K results (P$@$k) or mean average precision (mAP), the results does not take into account similarities among classes and varying misclassification costs. Hence, we choose the metric proposed in ~\cite{deng2011hierarchical}, average hierarchical precision $@$K (AHP@K) which is aware of the semantic relationships of the classes. In our experiments, we present results as AHP$@$40, because it indicates the average number of images per class in Indian heritage monument dataset. The training set and validation set is roughly divided in 40 images and 10 images respectively from each class. We use each of the test image as individual query, to retrieve semantically similar images from the remaining ones. Retrieval is performed by ranking the images in dataset in decreasing order of the dot product with the query image in the feature space. We evaluate semantic embeddings, learnt with combined cost function of correlation loss and cross-entropy loss, in addition with DeLF based re-ranking. As a baseline, the comparison is done with features extracted from the last layer of the same network architecture, trained with cross-entropy loss namely classification-based in Table \ref{tab:reter_mahp}. Further comparison is done with label embeddings networks \cite{sun2017label} as shown in table~\ref{tab:reter_mahp}. All methods have been experimented with identical network architectures and trained with same conditions. We also plot a graph of hierarchical precision against the top K correctly retrieved results with different CNN architectures as shown in Fig. \ref{fig:mahp40}. 

\begin{figure}[htbp]
    \centering
    \includegraphics[width=7cm]{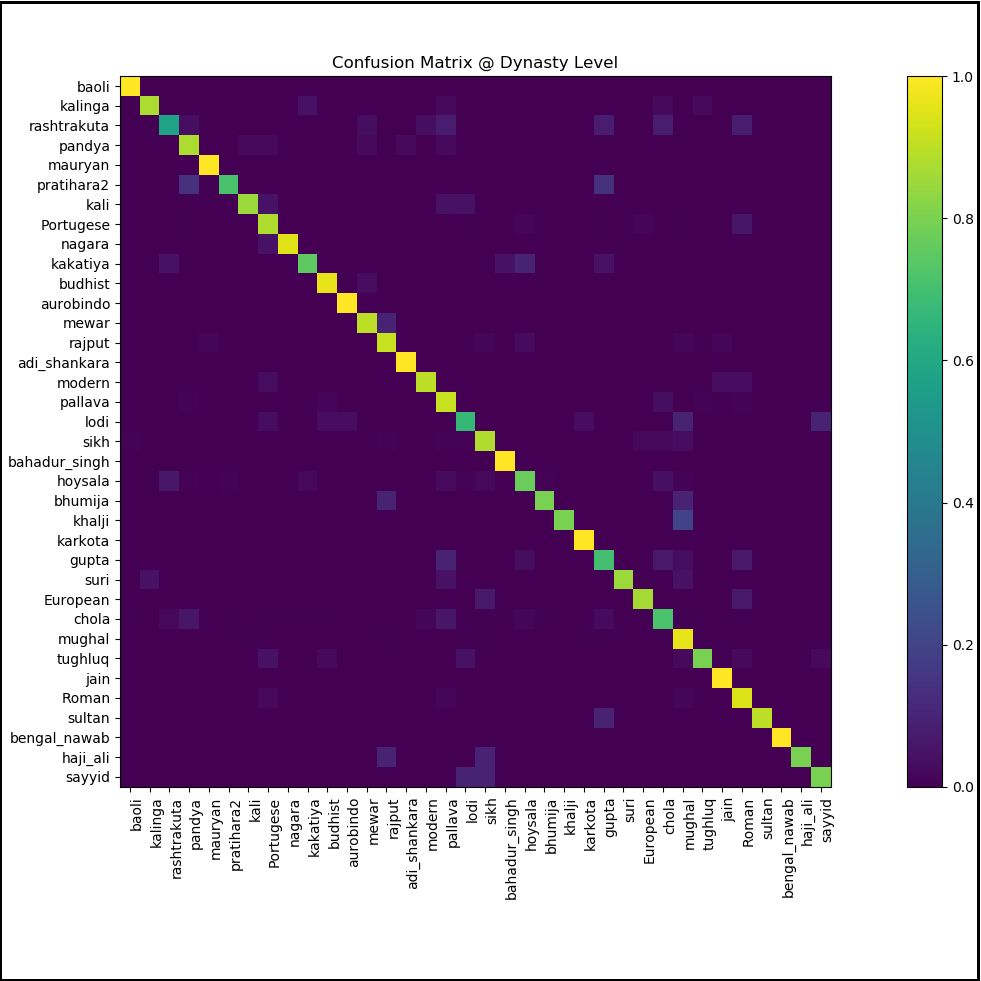}
    \caption{Confusion matrix for dynasty level classification with Semantic Embeddings from ResNet-50 finetuned network}
    \label{fig:conf_dyna_type}
\end{figure}

\begin{figure*}[hbtp]
\centering
  \fbox{
\includegraphics[scale=.5]{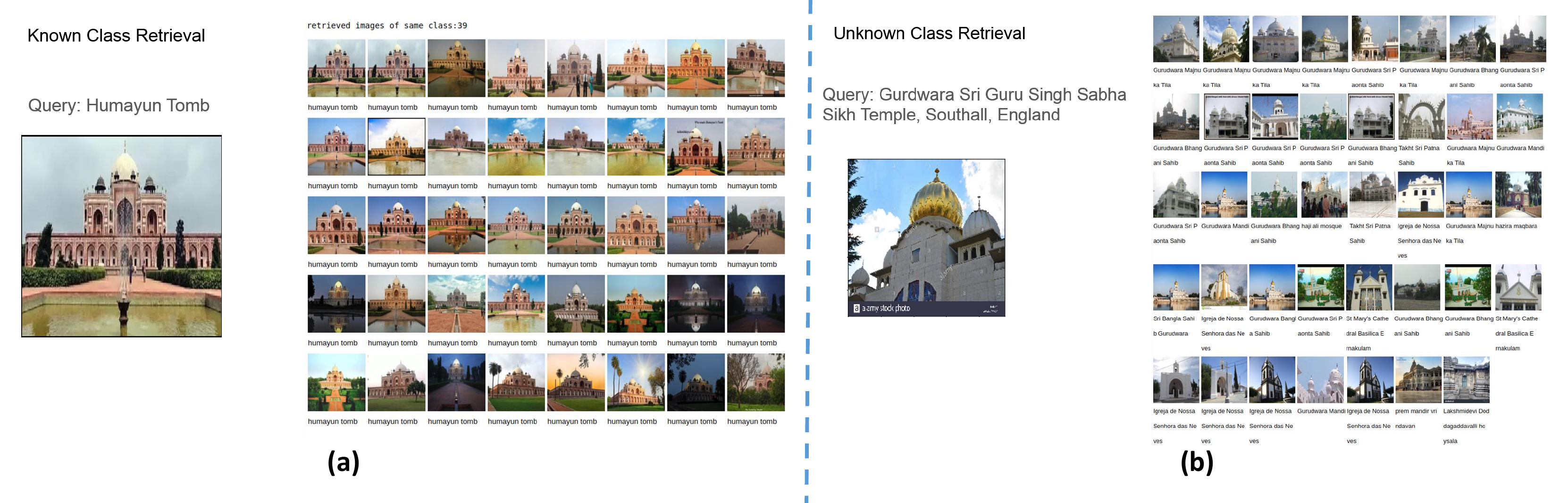}}
\caption{(a) Retrieval results for a monument with known class (b) Unknown (unseen) class retrieval.}
\label{fig:salprop}
\end{figure*}

\begin{figure}[hbtp]
\centering

$\begin{array}{cccc}
\includegraphics[width=2cm,height=2cm]{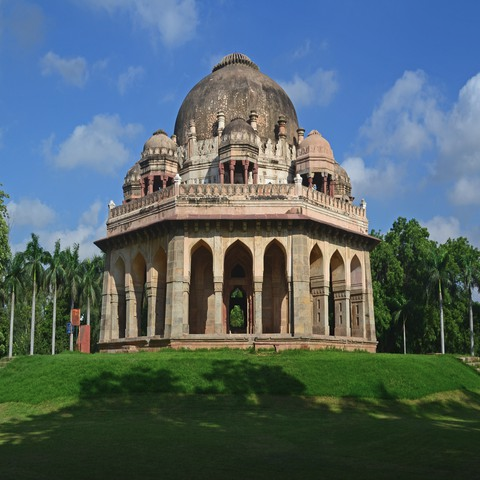} &
\includegraphics[width=2cm,height=2cm]{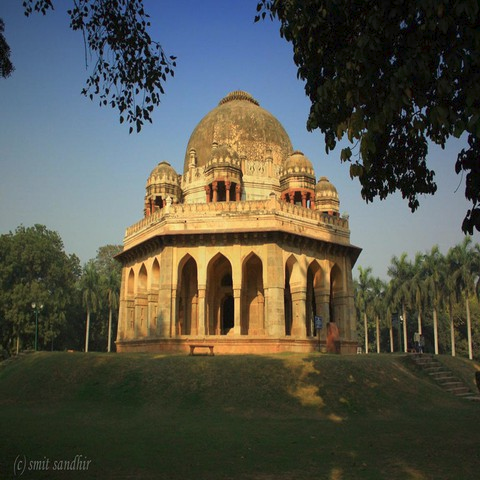} & 
\includegraphics[width=2cm,height=2cm]{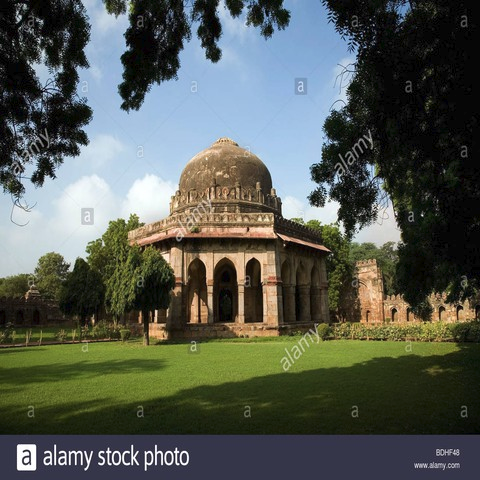} &
\includegraphics[width=2cm,height=2cm]{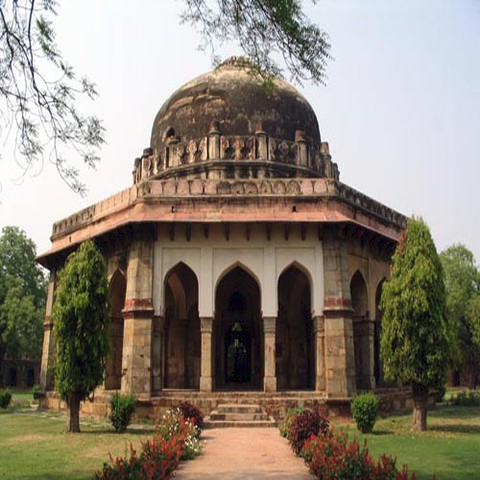}\\
(a) & (b) & (c) & (d)\\
\end{array}$
\caption{Confusion between Visually similar sub-classes (a)Sayyid Tomb classified as Lodi Tomb (b) Sayyid Tomb classified as Sayyid Tomb (c) Lodi Tomb classified as Sayyid Tomb (d) Lodi Tomb classified as Lodi Tomb}
\label{fig:confu}
\end{figure}

% The mean values of HP@k on Indian Monuments over all the queries at the first 40 values of k are reported in Table \ref{tab:reter_mahp}. Pre-trained ResNet-50 architecture over CIFAR-100 with semantic embeddings achieves higher hierarchical precision than other architectures with semantic embeddings.

Similarly for image classification, we evaluate the semantic embeddings on basis of accuracy obtained with softmax function applied to fully-connected layer on top of the embedding layer. Further classification accuracy is compared with label embedding network model \cite{sun2017label} after applying softmax function and the baseline classification model without any embedding layer as shown in table \ref{tab:classacc}. As shown in table \ref{tab:classacc} pre-trained ResNet-50 model(*) performs best in monument image classification on given dataset. The performance of semantic embeddings is lower because we consider Top-1 accuracy for classification.

We further perform experiments to improve the performance of pre-trained ResNet-50 model with different input image sizes and evaluate the corresponding inference speed as shown in table \ref{tab:inference}. We also construct confusion matrix for the best performing ResNet-50 model at different levels of image classification as shown in Fig. \ref{fig:conf_mon_type} and in Fig.\ref{fig:conf_dyna_type}. In Fig. \ref{fig:conf_mon_type} the confusion matrix for monument type level image classification is shown and Fig. \ref{fig:conf_dyna_type} shows the confusion matrix for dynasty level image classification. From these confusion matrices, we observe that proposed framework is not only able to categorize images in classes at higher level of hierarchy such as monument type level with considerable high accuracy but also categorize them in classes at lower level of hierarchy such as dynasty level with good accuracy. Usually the classification performance goes down when we descend from higher level to lower level of hierarchy due to high visual similarity between the categories at lower levels. Table\ref{tab:confused} shows the confusion matrix at leaf node level between 2 visually similar monument classes.

% In our dataset, each class of monument has approximately 50 images. The training set and validation set is roughly divided in 40 images and 10 images respectively from each class. We use each of the test image as individual query, to retrieve semantically similar images from the remaining ones. Retrieval is performed by ranking the images in dataset in decreasing order of the dot product with the query image in the feature space.

%%%%%%%%%%%%%%%%%%%%%%%%%%%%%%%%%%%%%%%%%%%%%%%%%%%%%%%%%%%%%%%%%%%%%%%%%%

\subsection{Qualitative Evaluation}
%\vspace{-0.2cm}

% The goal of finding similar images involves assessing the images on the basis of visual similarity which can be computed in terms of MSE, MAD, SSIM or feature based distance measures while finding images with similar semantics involves assessing the relationship of concepts present in the image such as in terms of taxonomy or hierarchy. 
% Here, retrieval of images while preserving semantics is relevant for us. Hence the visual features embeddings (DeLF) complements semantic feature embedding information and thus improving the overall image retrieval performance. 
% As opposed to standard image retrieval pipelines (which retrieves visually similar monuments independently from their class), here we address the issue of semantics infused image retrieval where the classes which belong to similar eras or constructed under common dynasties or belonging to same architectural styles will be retrieved i.e not similar monuments but of the same class will be outputted as retrieval results. The DeLF based reranking will finetune these retrieval results by further sorting the similar monuments belonging to the same era (class). 

Figure \ref{fig:salprop} shows the qualitative image retrieval results for known and unknown monument class, while training. The left side of the figure \ref{fig:salprop} shows the retrieved images for test query 'Humayun's Tomb' where 'Humayun's Tomb' class is present during the training. On the right side of the figure \ref{fig:salprop} shows the retrieved images for test query 'Gurudwara Sri Guru Singh Sabha Sikh Temple' from Southall England where the concerned monument class is not present in during the training. Hence the qualitative results shows that globally, semantic information is well captured in the semantic embeddings and re-ranking of retrieved images with DeLF features improved the results locally. It confirms that even though the trained model for given monument classes does not encompass all the monuments that exist, we can still retrieve similar looking monuments from the hierarchy at the level of architectural styles.

Figure \ref{fig:confu} shows the visual examples for the confusion matrix shown in table \ref{tab:confused}. It captures the misclassification at leaf node level in visually similar sub-classes, where 'Lodi Tomb' is classified as 'Sayyid Tomb' and vice versa. Here, we observe that in case of visually similar sub-classes there are very few local structures or subtle differences which distinguishes them from one another, thus leading to more confusion. 

%We observe that our semantic representation captures the visually similar global structures while 

\section{Conclusion}
\label{sec:conclusion}
In this paper, the proposed framework for semantic feature embeddings integrates the prior knowledge about the semantics between the classes with deep learning with the help of an Indian heritage monuments taxonomy based on era, dynasty and architecture styles. Thus retrieved images from the database using this semantic feature embeddings are not only visually similar but also semantically similar to the test query image. The exhaustive set of results confirms that global and local representation of monument images is very well captured by the semantic feature embeddings and the DeLF signatures.

\bibliographystyle{IEEEtran}
\bibliography{refs}

% Generated by IEEEtran.bst, version: 1.14 (2015/08/26)
\begin{thebibliography}{10}
\providecommand{\url}[1]{#1}
\csname url@samestyle\endcsname
\providecommand{\newblock}{\relax}
\providecommand{\bibinfo}[2]{#2}
\providecommand{\BIBentrySTDinterwordspacing}{\spaceskip=0pt\relax}
\providecommand{\BIBentryALTinterwordstretchfactor}{4}
\providecommand{\BIBentryALTinterwordspacing}{\spaceskip=\fontdimen2\font plus
\BIBentryALTinterwordstretchfactor\fontdimen3\font minus
  \fontdimen4\font\relax}
\providecommand{\BIBforeignlanguage}[2]{{%
\expandafter\ifx\csname l@#1\endcsname\relax
\typeout{** WARNING: IEEEtran.bst: No hyphenation pattern has been}%
\typeout{** loaded for the language `#1'. Using the pattern for}%
\typeout{** the default language instead.}%
\else
\language=\csname l@#1\endcsname
\fi
#2}}
\providecommand{\BIBdecl}{\relax}
\BIBdecl

\bibitem{goel2012buildings}
A.~Goel, M.~Juneja, and C.~Jawahar, ``Are buildings only instances? exploration
  in architectural style categories,'' in \emph{Proceedings of the Eighth
  Indian Conference on Computer Vision, Graphics and Image Processing}, 2012,
  pp. 1--8.

\bibitem{chaudhury2015multimedia}
S.~Chaudhury, A.~Mallik, and H.~Ghosh, \emph{Multimedia ontology:
  representation and applications}.\hskip 1em plus 0.5em minus 0.4em\relax CRC
  Press, 2015.

\bibitem{doersch2015makes}
C.~Doersch, S.~Singh, A.~Gupta, J.~Sivic, and A.~A. Efros, ``What makes paris
  look like paris?'' \emph{Communications of the ACM}, vol.~58, no.~12, pp.
  103--110, 2015.

\bibitem{chelaramani2017interactive}
S.~Chelaramani, V.~Muthireddy, and C.~Jawahar, ``An interactive tour guide for
  a heritage site,'' in \emph{Proceedings of the IEEE International Conference
  on Computer Vision Workshops}, 2017, pp. 2943--2952.

\bibitem{shukla2017computer}
P.~Shukla, B.~Rautela, and A.~Mittal, ``A computer vision framework for
  automatic description of indian monuments,'' in \emph{2017 13th International
  Conference on Signal-Image Technology \& Internet-Based Systems
  (SITIS)}.\hskip 1em plus 0.5em minus 0.4em\relax IEEE, 2017, pp. 116--122.

\bibitem{radenovic2018fine}
F.~Radenovi{\'c}, G.~Tolias, and O.~Chum, ``Fine-tuning cnn image retrieval
  with no human annotation,'' \emph{IEEE transactions on pattern analysis and
  machine intelligence}, 2018.

\bibitem{xu2018cross}
P.~Xu, Q.~Yin, Y.~Huang, Y.-Z. Song, Z.~Ma, L.~Wang, T.~Xiang, W.~B. Kleijn,
  and J.~Guo, ``Cross-modal subspace learning for fine-grained sketch-based
  image retrieval,'' \emph{Neurocomputing}, vol. 278, pp. 75--86, 2018.

\bibitem{song2018binary}
J.~Song, T.~He, L.~Gao, X.~Xu, A.~Hanjalic, and H.~T. Shen, ``Binary generative
  adversarial networks for image retrieval,'' in \emph{Thirty-Second AAAI
  Conference on Artificial Intelligence}, 2018.

\bibitem{zhao2015deep}
F.~Zhao, Y.~Huang, L.~Wang, and T.~Tan, ``Deep semantic ranking based hashing
  for multi-label image retrieval,'' in \emph{Proceedings of the IEEE
  conference on computer vision and pattern recognition}, 2015, pp. 1556--1564.

\bibitem{lu2016latent}
X.~Lu, X.~Zheng, and X.~Li, ``Latent semantic minimal hashing for image
  retrieval,'' \emph{IEEE Transactions on Image Processing}, vol.~26, no.~1,
  pp. 355--368, 2016.

\bibitem{sivic2003video}
J.~Sivic and A.~Zisserman, ``Video google: A text retrieval approach to object
  matching in videos,'' in \emph{null}.\hskip 1em plus 0.5em minus 0.4em\relax
  IEEE, 2003, p. 1470.

\bibitem{sanchez2013image}
J.~S{\'a}nchez, F.~Perronnin, T.~Mensink, and J.~Verbeek, ``Image
  classification with the fisher vector: Theory and practice,''
  \emph{International journal of computer vision}, vol. 105, no.~3, pp.
  222--245, 2013.

\bibitem{philbin2007object}
J.~Philbin, O.~Chum, M.~Isard, J.~Sivic, and A.~Zisserman, ``Object retrieval
  with large vocabularies and fast spatial matching,'' in \emph{2007 IEEE
  Conference on Computer Vision and Pattern Recognition}.\hskip 1em plus 0.5em
  minus 0.4em\relax IEEE, 2007, pp. 1--8.

\bibitem{krizhevsky2012imagenet}
A.~Krizhevsky, I.~Sutskever, and G.~E. Hinton, ``Imagenet classification with
  deep convolutional neural networks,'' in \emph{Advances in neural information
  processing systems}, 2012, pp. 1097--1105.

\bibitem{lin2015deep}
K.~Lin, H.-F. Yang, J.-H. Hsiao, and C.-S. Chen, ``Deep learning of binary hash
  codes for fast image retrieval,'' in \emph{Proceedings of the IEEE conference
  on computer vision and pattern recognition workshops}, 2015, pp. 27--35.

\bibitem{noh2017large}
H.~Noh, A.~Araujo, J.~Sim, T.~Weyand, and B.~Han, ``Large-scale image retrieval
  with attentive deep local features,'' in \emph{Proceedings of the IEEE
  International Conference on Computer Vision}, 2017, pp. 3456--3465.

\bibitem{johnson2015image}
J.~Johnson, R.~Krishna, M.~Stark, L.-J. Li, D.~Shamma, M.~Bernstein, and
  L.~Fei-Fei, ``Image retrieval using scene graphs,'' in \emph{Proceedings of
  the IEEE conference on computer vision and pattern recognition}, 2015, pp.
  3668--3678.

\bibitem{csurka2004visual}
G.~Csurka, C.~Dance, L.~Fan, J.~Willamowski, and C.~Bray, ``Visual
  categorization with bags of keypoints,'' in \emph{Workshop on statistical
  learning in computer vision, ECCV}, vol.~1, no. 1-22.\hskip 1em plus 0.5em
  minus 0.4em\relax Prague, 2004, pp. 1--2.

\bibitem{rublee2011orb}
E.~Rublee, V.~Rabaud, K.~Konolige, and G.~R. Bradski, ``Orb: An efficient
  alternative to sift or surf.'' in \emph{ICCV}, vol.~11, no.~1.\hskip 1em plus
  0.5em minus 0.4em\relax Citeseer, 2011, p.~2.

\bibitem{arandjelovic2012three}
R.~Arandjelovi{\'c} and A.~Zisserman, ``Three things everyone should know to
  improve object retrieval,'' in \emph{2012 IEEE Conference on Computer Vision
  and Pattern Recognition}.\hskip 1em plus 0.5em minus 0.4em\relax IEEE, 2012,
  pp. 2911--2918.

\bibitem{delhumeau2013revisiting}
J.~Delhumeau, P.-H. Gosselin, H.~J{\'e}gou, and P.~P{\'e}rez, ``Revisiting the
  vlad image representation,'' in \emph{Proceedings of the 21st ACM
  international conference on Multimedia}.\hskip 1em plus 0.5em minus
  0.4em\relax ACM, 2013, pp. 653--656.

\bibitem{xia2016privacy}
Z.~Xia, X.~Wang, L.~Zhang, Z.~Qin, X.~Sun, and K.~Ren, ``A privacy-preserving
  and copy-deterrence content-based image retrieval scheme in cloud
  computing,'' \emph{IEEE transactions on information forensics and security},
  vol.~11, no.~11, pp. 2594--2608, 2016.

\bibitem{zhu2016unsupervised}
L.~Zhu, J.~Shen, L.~Xie, and Z.~Cheng, ``Unsupervised visual hashing with
  semantic assistant for content-based image retrieval,'' \emph{IEEE
  Transactions on Knowledge and Data Engineering}, vol.~29, no.~2, pp.
  472--486, 2016.

\bibitem{piras2017information}
L.~Piras and G.~Giacinto, ``Information fusion in content based image
  retrieval: A comprehensive overview,'' \emph{Information Fusion}, vol.~37,
  pp. 50--60, 2017.

\bibitem{revaud2019learning}
J.~Revaud, J.~Almaz{\'a}n, R.~S. Rezende, and C.~R.~d. Souza, ``Learning with
  average precision: Training image retrieval with a listwise loss,'' in
  \emph{Proceedings of the IEEE International Conference on Computer Vision},
  2019, pp. 5107--5116.

\bibitem{babenko2015aggregating}
A.~Babenko and V.~Lempitsky, ``Aggregating local deep features for image
  retrieval,'' in \emph{Proceedings of the IEEE international conference on
  computer vision}, 2015, pp. 1269--1277.

\bibitem{gordo2016deep}
A.~Gordo, J.~Almaz{\'a}n, J.~Revaud, and D.~Larlus, ``Deep image retrieval:
  Learning global representations for image search,'' in \emph{European
  conference on computer vision}.\hskip 1em plus 0.5em minus 0.4em\relax
  Springer, 2016, pp. 241--257.

\bibitem{gordo2017end}
A.~Gordo, J.~Almazan, J.~Revaud, and D.~Larlus, ``End-to-end learning of deep
  visual representations for image retrieval,'' \emph{International Journal of
  Computer Vision}, vol. 124, no.~2, pp. 237--254, 2017.

\bibitem{salvador2016faster}
A.~Salvador, X.~Gir{\'o}-i Nieto, F.~Marqu{\'e}s, and S.~Satoh, ``Faster r-cnn
  features for instance search,'' in \emph{Proceedings of the IEEE Conference
  on Computer Vision and Pattern Recognition Workshops}, 2016, pp. 9--16.

\bibitem{tolias2015particular}
G.~Tolias, R.~Sicre, and H.~J{\'e}gou, ``Particular object retrieval with
  integral max-pooling of cnn activations,'' \emph{arXiv preprint
  arXiv:1511.05879}, 2015.

\bibitem{altwaijry2016learning}
H.~Altwaijry, E.~Trulls, J.~Hays, P.~Fua, and S.~Belongie, ``Learning to match
  aerial images with deep attentive architectures,'' in \emph{Proceedings of
  the IEEE Conference on Computer Vision and Pattern Recognition}, 2016, pp.
  3539--3547.

\bibitem{teichmann2018detect}
M.~Teichmann, A.~Araujo, M.~Zhu, and J.~Sim, ``Detect-to-retrieve: Efficient
  regional aggregation for image search,'' \emph{arXiv preprint
  arXiv:1812.01584}, 2018.

\bibitem{kim2018regional}
J.~Kim and S.-E. Yoon, ``Regional attention based deep feature for image
  retrieval.'' in \emph{BMVC}, 2018, p. 209.

\bibitem{barz2019hierarchy}
B.~Barz and J.~Denzler, ``Hierarchy-based image embeddings for semantic image
  retrieval,'' in \emph{2019 IEEE Winter Conference on Applications of Computer
  Vision (WACV)}.\hskip 1em plus 0.5em minus 0.4em\relax IEEE, 2019, pp.
  638--647.

\bibitem{zhao2011large}
B.~Zhao, F.~Li, and E.~P. Xing, ``Large-scale category structure aware image
  categorization,'' in \emph{Advances in Neural Information Processing
  Systems}, 2011, pp. 1251--1259.

\bibitem{chang2015large}
J.~Y. Chang and K.~M. Lee, ``Large margin learning of hierarchical semantic
  similarity for image classification,'' \emph{Computer Vision and Image
  Understanding}, vol. 132, pp. 3--11, 2015.

\bibitem{he2016deep}
K.~He, X.~Zhang, S.~Ren, and J.~Sun, ``Deep residual learning for image
  recognition,'' in \emph{Proceedings of the IEEE conference on computer vision
  and pattern recognition}, 2016, pp. 770--778.

\bibitem{russakovsky2015imagenet}
O.~Russakovsky, J.~Deng, H.~Su, J.~Krause, S.~Satheesh, S.~Ma, Z.~Huang,
  A.~Karpathy, A.~Khosla, M.~Bernstein \emph{et~al.}, ``Imagenet large scale
  visual recognition challenge,'' \emph{International journal of computer
  vision}, vol. 115, no.~3, pp. 211--252, 2015.

\bibitem{deng2011hierarchical}
J.~Deng, A.~C. Berg, and L.~Fei-Fei, ``Hierarchical semantic indexing for large
  scale image retrieval,'' in \emph{CVPR 2011}.\hskip 1em plus 0.5em minus
  0.4em\relax IEEE, 2011, pp. 785--792.

\bibitem{loshchilov2016sgdr}
I.~Loshchilov and F.~Hutter, ``Sgdr: Stochastic gradient descent with warm
  restarts,'' \emph{arXiv preprint arXiv:1608.03983}, 2016.

\bibitem{kumar2018improving}
A.~Kumar, S.~Bhowmick, N.~Jayanthi, and S.~Indu, ``Improving landmark
  recognition using saliency detection and feature classification,''
  \emph{arXiv preprint arXiv:1811.12748}, 2018.

\bibitem{podder2018ontology}
D.~Podder, J.~Mukherjee, S.~M. Aswatha, J.~Mukherjee, and S.~Sural,
  ``Ontology-driven content-based retrieval of heritage images,'' in
  \emph{Heritage Preservation}.\hskip 1em plus 0.5em minus 0.4em\relax
  Springer, 2018, pp. 143--160.

\bibitem{sun2017label}
X.~Sun, B.~Wei, X.~Ren, and S.~Ma, ``Label embedding network: Learning label
  representation for soft training of deep networks,'' \emph{arXiv preprint
  arXiv:1710.10393}, 2017.

\end{thebibliography}

\end{document}